\pdfoutput=1

\documentclass[11pt]{article}

\usepackage[final]{acl}

\usepackage{times}
\usepackage{latexsym}
\usepackage{microtype}
\usepackage[utf8]{inputenc}
\usepackage[T1]{fontenc}

\usepackage{graphicx}
\usepackage{stfloats}
\usepackage{inconsolata}
\usepackage{soul} 
\usepackage{xcolor} 
\usepackage{subcaption}
\usepackage{booktabs}
\usepackage{multirow}
\usepackage{adjustbox}
\usepackage{amsmath}
\usepackage{calligra}
\usepackage{float}
\usepackage{hyperref}
\usepackage{tabularx}
\usepackage{array}

\usepackage{xspace}
\newcommand{\psychemb}{\text{PsychEmb}\xspace}

\title{WhiSPA: Semantically and Psychologically Aligned Whisper with Self-Supervised Contrastive and Student-Teacher Learning}

\author{
    \textbf{Rajath Rao\textsuperscript{1}}, 
    \textbf{Adithya V Ganesan\textsuperscript{1}}, 
    \textbf{Oscar Kjell\textsuperscript{1}}, 
    \textbf{Jonah Luby\textsuperscript{1}}, 
    \textbf{Akshay Raghavan\textsuperscript{1}}, 
    \\
    \textbf{Scott Feltman\textsuperscript{1}} 
    \textbf{Whitney Ringwald\textsuperscript{2}}, 
    \textbf{Ryan L. Boyd\textsuperscript{3}}, 
    \textbf{Benjamin Luft\textsuperscript{1}}, 
    \\
    \textbf{Camilo Ruggero\textsuperscript{3}}, 
    \textbf{Neville Ryant\textsuperscript{4}}, 
    \textbf{Roman Kotov\textsuperscript{1}}, 
    \textbf{H. Andrew Schwartz\textsuperscript{1}} 
    \\
    \textsuperscript{1}Stony Brook University, 
    \textsuperscript{2}University of Minnesota\\
    \textsuperscript{3}University of Texas at Dallas, 
    \textsuperscript{4}University of Pennsylvania\\
    \small{
        \textbf{Correspondence:} \href{mailto:email@domain}{rajath.rao@stonybrook.edu}, \href{mailto:email@domain}{has@cs.stonybrook.edu}
    }
}
\begin{document}
\maketitle

\begin{abstract}
Current speech encoding pipelines often rely on an additional text-based LM to get robust representations of human communication, even though SotA speech-to-text models often have a LM within.
This work proposes an approach to improve the LM within an audio model such that the subsequent text-LM is unnecessary. 
We introduce \textbf{WhiSPA} (\textbf{Whi}sper with \textbf{S}emantic and \textbf{P}sychological \textbf{A}lignment), which leverages a novel audio training objective: contrastive loss with a language model embedding as a teacher.
Using over 500k speech segments from mental health audio interviews, we evaluate the utility of aligning Whisper’s latent space with semantic representations from a text autoencoder (SBERT) and lexically derived embeddings of basic psychological dimensions: emotion and personality.
Over self-supervised affective tasks and downstream psychological tasks, WhiSPA surpasses current speech encoders, achieving an average error reduction of $73.4\%$ and $83.8\%$, respectively. 
WhiSPA demonstrates that it is not always necessary to run a subsequent text LM on speech-to-text output in order to get a rich psychological representation of human communication.
\end{abstract}

\section{Introduction}

\begin{figure}[!th]
\includegraphics[width=\columnwidth]{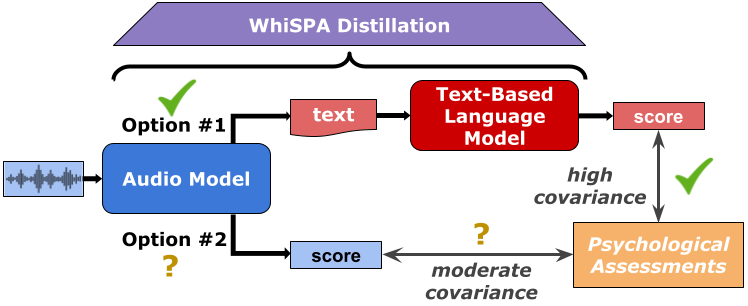}
\caption{Speech processing pipelines that are further processed by language models (option \#1 above) often yield higher accuracies than those produced solely by SotA audio models (option \#2).
Our distillation introduces a speech encoder which streamlines the pipeline and exhibits similar performance to a text-based LM.}
\label{fig:spirit_diagram}
\end{figure}

Human communication is inherently multimodal, but AI integration of modalities is often fragmented~\cite{lazaro2021multimodal, gu2017-speechtextsep}. Speech to text models, like Whisper~\cite{radford2022whisper}, are often pipelined into text-based language models (LMs)~\cite{chuang2020speechbert} in order to get the most accurate speech-based representations (see \autoref{fig:spirit_diagram}). 
This often results in redundant computational costs from having two LMs in the pipeline (one within the audio model and one for the text LM) and representations remain incomplete of the full spectrum of human expressions \cite{zhang2023bridgingemotionalsemanticgap, lian2023surveyspeechtextface}.
This is especially important for psychological and social scientific applications where representations from text-based LMs demonstrate superior performance than direct speech representations~\cite{Lukac2024-tb, chen2024textbetterspeech}.


Here, we seek to bridge the \textit{semantic and psychological representation gap} between speech-based LMs present in audio models and text-based LMs.
We introduce a speech encoding model, \textbf{WhiSPA}\footnote{\url{https://github.com/humanlab/WhiSPA}} (\textbf{Whi}sper with \textbf{S}emantic and \textbf{P}sychological \textbf{A}lignment), which aligns a pre-trained speech recognition model, Whisper \cite{radford2022whisper}, with the latent dimensions from SBERT \cite{reimers2019sbert}, intended to better capture semantics and deeper psychological information~\cite{v-ganesan-etal-2022-wwbp, park2014personality}.
Such alignment reduces computational and memory inefficiencies, circumventing the need for a second text encoder, as it enables a unified cross-modal representation between speech and language models.
Still, since text is derivable from speech, speech should intrinsically be mappable to the same rich semantic features from the text.

Our focus on psychological or human-level tasks reflects a growing demand for foundation models to better understand the intrinsic qualities of human communication~\cite{soni-etal-2024-large}.
As ~\citet{clark1992askingquestions} put it, \textit{``The common misconception is that language has to do with words and what they mean. It does not. It has to do with people and what they mean.''}
and specifically how well the representations capture information \textit{about the people} communicating~\cite{hovy2021importance, soni2022hart}.
More specifically, psychological studies have suggested mental health attributes are highly multimodal as they are influenced by subtle nuances in voice and content~\cite{sartori2023psychlm, chen2024textbetterspeech}.

Our \textbf{main contributions} include:
(1) The development of \textbf{WhiSPA} (\textbf{Whi}sper with \textbf{S}emantic and \textbf{P}sychological \textbf{A}lignment), with a novel alignment objective, 
(2) Evaluation of the hypothesis that aligning text and audio latent spaces can significantly enhance audio-based representations for a deeper semantic and psychological understanding of human communication, 
(3) Demonstration of significant accuracy improvements in self-supervised tasks and downstream psychological tasks over systematically tested variants of WhiSPA. 
We find that: 
(a) aligning with text-based semantic and psychological representations drastically improves audio representations, including SotA person-level psychological assessments;
(b) a Noise Contrastive Estimation loss yielded a more optimal convergence in aligning Whisper's latent space with semantic and psychological dimensions.
and (c) for downstream psychological tasks, there was almost no benefit in utilizing SBERT representations on top of WhiSPA's, suggesting the same information from a text LM can be captured with the LM of the audio model and thus it is not necessary to pipeline another text LM after the audio model. 


\section{Background}
This work builds on top of Whisper \cite{radford2022whisper}, OpenAI’s SotA automatic speech recognition (ASR) foundation model. 
We chose Whisper over other alternatives such as HuBERT and Wav2Vec2-BERT, since previous works~\cite{kyung2024enhancing, yang2023whispersota} have shown that Whisper has a stronger language encoding module at capturing speaker attributes due to its pretraining objective of transcription/translation.

Recent advances in foundational speech technologies, like Whisper and HuBERT, have vastly improved the performances on speech recognition tasks~\cite{radford2022whisper, hsu2021hubert}.
However, they have limited ability to capture deeper semantics and speaker attributes compared to a text-based language model~\cite{chen2024textbetterspeech, dong2022improving}. 
Prior works that have addressed this have targeted a very narrow scope of psychological attributes~\cite{busso2008-iemocap}.
These gaps underscore the need for methodologies that bridge speech encoders’ acoustic robustness with the psychological depth of text-based language models---a challenge we address by embedding fundamental psychological dimensions present in one's speech.

Multi-level fusion architectures leveraging both acoustic and lexical features have shown to improve performance in emotion recognition, speaker identification, and other downstream tasks.
For instance, \cite{zhao2022multilevelfusionwav2vec20} demonstrates that coattention-based early fusion and late fusion using Wav2Vec2.0 \cite{baevski2020wav2vec2, schneider2019wav2vec} and BERT \cite{devlin2019bert} outperform SotA emotion recognition benchmarks.
Other recent works inject acoustic nuances into language models using textual descriptions of speech characteristics \cite{wu2024silentletters} or common-sense reasoning through historical utterances from the speaker \cite{fu2024ckercjointlarge}.
However, this approach does not fully leverage the cross-modal dependencies between text and audio, as it remains unimodal, relying solely on textual inputs rather than raw acoustic representations.

Prior works in cross-modal alignment provide foundational insights for this integration. 
Compositional Contrastive Learning~\cite{chen2021distilling} distilled audio-visual knowledge into video representations by aligning teacher-student embeddings across modalities, embedding rich semantics from teacher-audio and image models into the student-video model.
In another work, ~\citet{dong2022improving} improved the accuracy of intent classification of spoken language by employing a contrastive loss using both speech and language features. 
These works highlight that the cross-modal alignment objective embeds information from different modalities into shared spaces to capture their relationships, while contrastive learning aids in grouping related inputs across different modalities (e.g., audio and text segments) while separating unrelated pairs \cite{ye2022crossmodalspeech}.
Efforts to align text and audio include SpeechBERT \cite{chuang2020speechbert}, which adapted BERT’s framework~\cite{devlin2019bert} to paired speech-text data, and SLAM (Speech-Language Aligned Models) \cite{bapna2022slam}, which optimized joint embedding spaces to improve downstream tasks like speech recognition and audio-text retrieval.
To the best of our knowledge, this is the first work to perform cross-modal learning to endow the foundational speech model with richer semantic and psychological representations. 

\section{Data \& Tasks}
\paragraph{Audio Datasets.}
We utilize two psychological, mental health-focused datasets for training and evaluation: \textbf{WTC-Segments} (WTC) \cite{kjell2024wtc} and \textbf{HiTOP-Segments} (HiTOP) \cite{kotov2022-hitop}.
WTC recordings were completed by patients in a clinic for World Trade Center (9/11) responders who came for a health monitoring visit.
HiTOP interviews were completed by outpatients with psychiatric diagnoses who were recruited by the study team to complete a research interview. Both datasets consist of paired audio-text data, ensuring alignment between spoken content and its corresponding textual transcription.

\begin{figure*}[b!]
\centering
\includegraphics[width=0.8\textwidth]{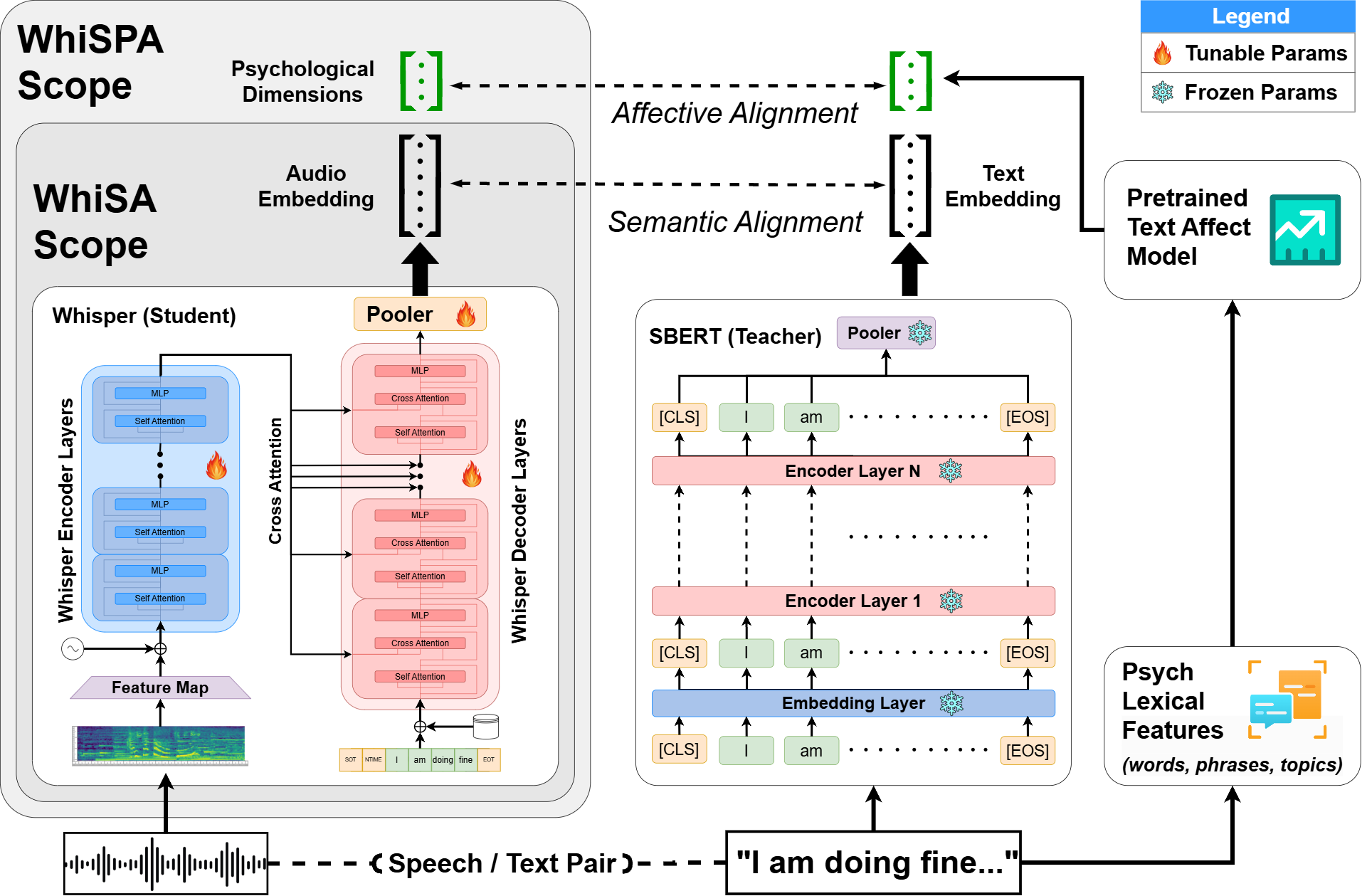}
\caption{Diagram of WhiSA and WhiSPA training procedure involving a student-teacher model paradigm. Whisper (left) is semantically aligned to the ground truth embeddings encoded by SBERT (right). When PsychEmb features are included in the alignment function, the WhiSPA framework semantically and psychologically aligns the corresponding embeddings with contrastive loss criteria.}
\label{fig:whispa_training}
\end{figure*}

\begin{table}[h!]
\centering
\small
\begin{tabular}{lcc}
\toprule
\textbf{Dataset} & \textbf{WTC} & \textbf{HiTOP} \\
\midrule
Total Segment Duration (\textit{hr}) & $\sim$252 & $\sim$474\\
Mean Segment Duration (\textit{s}) & 5.86  & 2.99\\
Total Audio Segments & 154,586 & 571,420\\
Total Participants & 1,396 & 524\\
\bottomrule
\end{tabular}
\caption{Audio dataset metadata (after preprocessing and filtering for participant-only speech).
\label{tab:dataset_table}}
\end{table}

From its source, WTC was curated from $\sim$6 minute interview recordings, on average, of patients responding to both personal and general questions in a structured manner \cite{kjell2024wtc}. 
Contrarily, HiTOP followed a semi-structured format, where patients described experiences on set topics while also organically conversing with the interviewer. Once filtered for audio segments solely spoken by patients, interviews generally ranged from 45 to 90 minutes, yielding a voluminous and broadened set of audio segments \cite{kotov2022-hitop}. The recordings were diarized using NVIDIA NeMo and transcribed with \href{https://huggingface.co/openai/whisper-large-v2}{\texttt{whisper-large-v2}}.

\paragraph{Psychological Assessments.} For each dataset, psychological measures were collected for each user. For WTC, each subject completed the self-reported PTSD CheckList (PCL), yielding scores for four specific subscales: Re-experiencing (REX), Avoidance (AVO), Negative Alterations in Mood (NAM), Hyperarousal (HYP). For HiTOP, trained interviewers provided ratings for the following six psychopathology scales: Internalizing (INT), Disinhibition (DIS), Antagonism (ANT), Somatoform (SOM), Thought-Disorder (THD), and Detachment (DET) \cite{kotov2022-hitop, kotov2024-hitop}.

To evaluate the encoding ability of WhiSPA for any given audio segment, we manually annotate a small subset from both datasets for valence and arousal dimensions expressed in their speech. 
Three random audio segments containing more than 5 uttered words from each user were sampled from each dataset and were annotated by two individuals with a background in psychology using the affective circumplex scale (\autoref{fig:affective-circumplex}).
This resulted in 300 audio segments, equally split between the two datasets. 

\paragraph{Self-Supervised \psychemb.} For each audio/text pair in our datasets, we extract theoretically derived psychological features using pre-trained lexica \cite{v-ganesan-etal-2022-wwbp}, which we refer to as \psychemb.
\psychemb broadly covers three domains of psychological constructs measured at different temporal granularity: (a) states, which reflect the emotional condition of the person at a point in time; (b) dispositions, which are slightly more stable than states and reflect the tendencies of humans to behave in certain ways and finally (c) the traits, which are long term stable characteristics \cite{park2014personality}. 
The ten dimensions of \psychemb are Valence (VAL), Arousal (ARO), Openness (OPE), Consciousness (CON), Extraversion (EXT), Agreeableness (AGR), Neuroticism (NEU), Anger (ANG), Anxiety (ANX), and Depression (DEP), each represented with scalar values.
After extracting self-supervised \psychemb dimensions for each segment across both datasets, we perform a 80:10:10 (train/val/test) split.

\section{Methodology}
Aligning audio representations directly with a text-based language model allows us to infuse the audio model's latent space with the rich semantic and affective details typically provided by text representations, thereby eliminating the need for a separate text LM. 
While this approach does not explicitly leverage the unique acoustic features of speech, it prioritizes efficiency by avoiding redundant processing and consistently delivers a semantically enriched representation—an advantage that is particularly critical for psychological and social scientific applications~\cite{Lukac2024-tb, chen2024textbetterspeech}.

\paragraph{Model Architecture.} 
We begin with the Whisper\footnote{Whisper-384 version: \href{https://huggingface.co/openai/whisper-tiny}{\texttt{whisper-tiny}}} encoder-decoder backbone~\cite{radford2022whisper}, which does not run autoregressively.
During training, audio segments are previously transcribed with \href{https://huggingface.co/openai/whisper-large-v2}{\texttt{whisper-large-v2}}, making it entirely self-supervised.
Likewise, SBERT and \psychemb representations were encoded using these transcriptions.
As seen in the Whisper (Student) portion of \autoref{fig:whispa_training}, we apply a mean pooling layer to the last hidden state of Whisper’s decoder yielding a singular representation for the input audio.
This representation is then pooled using a learnable dense layer, and the output serves as our embedding during alignment.
This aggregated representation is aligned to the pooled representations from pre-trained SBERT for semantic alignment and the \psychemb's dimensions for psychological alignment.
Throughout this paper, we denote the pre-trained Whisper model as \textbf{Whisper-384}, where the numeric suffix refers to the embedding dimensionality.

\subsection{Alignment Objective}
While fusion architectures focus on merging acoustic-textual features throughout layers, \textit{we contrast this paradigm} by directly aligning cross-modal latent spaces for deeper semantic and psychological representations from audio, bypassing the need for task-specific fusion architectures.
Our alignment objective aims to improve the semantic and psychological information encoded in Whisper (student) with the help of the representations from a strong text encoding teacher model like SBERT\footnote{SBERT-384 version: \href{https://huggingface.co/sentence-transformers/all-MiniLM-L12-v2}{\texttt{all-MiniLM-L12-v2}}} and \psychemb.
In this work, we explore two suitable candidate objective functions to align speech representations with text, which are described below in detail.    

\subsubsection{Cosine Similarity Loss \textit{(CS)}}

The success of the cosine similarity-based approach in building geometrically robust representations in SBERT motivated its use as an alignment objective in this work. 
We apply cosine similarity loss to the pooled audio embeddings and pooled SBERT embeddings, given by the following equation: 

\begin{equation}
\mathcal{L}^{CS}=\sum_{i \in \mathcal{I}}\mathcal{L}_{i}^{CS}\label{cs-loss}
\end{equation}
$$\mathcal{L}_{i}^{CS}=1-\text{sim}(\mathbf{A}_i, \mathbf{T}_i)$$
$$\text{sim}(\mathbf{A}_i, \mathbf{T}_i)=\frac{\mathbf{A}_i \cdot \mathbf{T}_i}{||\mathbf{A}_i|| \text{ }||\mathbf{T}_i||}$$

where $i \in \mathcal{I} \equiv \{1...N\}$ refers to the index of audio/text pair in a batch of N samples. 
$\mathbf{A}_i$ refers to the source audio embedding, $\mathbf{T}_i$ refers to its corresponding target text embedding, and sim() computes the cosine similarity between audio and text embeddings which produces a scalar value between $[-1, 1]$.
This loss can also be interpreted as the cosine diversity of the two embeddings.
To align the embedding spaces, we aim to maximize the cosine similarity between corresponding embedding pairs \cite{reimers2019sbert, sanh2020distilbert}, and hence decrease $\mathcal{L}^{CS}$. 

\subsubsection{Noise Contrastive Estimation Loss \textit{(NCE)}}

The Noise Contrastive Loss (\autoref{nce-loss}) is optimized to increase the cosine similarity between a pair of audio embedding and its corresponding text embedding while simultaneously increasing the differentiation between the audio embedding and randomly sampled text embeddings in that batch~\cite{ye2022crossmodalspeech}.

\begin{equation}
\mathcal{L}^{NCE}=\sum_{i \in \mathcal{I}}\mathcal{L}_{i}^{NCE}\label{nce-loss}
\end{equation}
$$\mathcal{L}_{i}^{NCE}=-\log\frac{\exp(\text{sim}(\mathbf{A}_i, \mathbf{T}_i) / \tau)}
{\sum_{b\in B(i)}\exp(\text{sim}(\mathbf{A}_i, \mathbf{T}_b) / \tau)}$$

where $\mathcal{L}_{i}^{NCE}$ refers to contrastive loss criteria in which pairwise cosine similarities are calculated for each audio embedding with all text embeddings in that batch.
Hence, there is only one positive text embedding that pairs with an audio embedding, while the remaining text embeddings from the batch serve as contrastive samples. 
Let $B(i) \in \mathcal{I}$, where $B(i)$ represents all other SBERT text embeddings in the batch such that $\mathbf{T}_b \neq \mathbf{T}_i$ \cite{ye2022crossmodalspeech, chen2020simclr, khosla2021scl}.
The variable $\mathbf{T}_b$ denotes the index of an arbitrary, negative SBERT text embedding sample and $\tau$, temperature, represents a tunable scalar parameter which is set to $0.1$.

\subsection{Whisper Semantically Aligned (WhiSA-384)}
WhiSA leverages a student-teacher model paradigm \cite{hinton2015distillingknowledgeneuralnetwork, sanh2020distilbert} to align Whisper's audio-based embeddings with SBERT's text-based embeddings, which serve as the teacher model. 
SBERT encodes corresponding text sentences into semantically rich embedding vectors, which serve as $\mathbf{T}$ in the above equations during training. 
Whisper's embeddings ($\mathbf{A}$ in the above equations), derived from its decoder's last hidden state, are aligned to these SBERT embeddings using the loss functions described above. 
This process is aimed at WhiSA to learn robust semantic representations directly from audio inputs by minimizing the cosine distance between Whisper and SBERT embeddings as shown in \autoref{fig:whispa_training}.

\subsection{Adding Psychological Alignment (WhiSPA)}
WhiSPA extends the WhiSA framework by augmenting \psychemb dimensions into Whisper's.
While maintaining the semantic alignment objective, WhiSPA injects the \psychemb dimensions into the SBERT embeddings under two settings:
(1) \textbf{with replacement}: We adopted a naive strategy of replacing the first ten dimensions of SBERT's embedding with the \psychemb dimensions to maintain the same number of latent dimensions between both models. We use \textbf{WhiSPA-384$_r$} to refer to this. (2) \textbf{with projection}: We concatenate the \psychemb dimensions to the text embedding from SBERT.
Consequently, this requires a $384 \times 10$ learnable projection matrix, $P$, to transform Whisper embeddings of dimensionality $384$ to $394$, which is then passed through a $TanH$ activation. This model goes by the name \textbf{WhiSPA-394}.
To address the numerical instability issues from modeling the \psychemb dimensions in its absolute range, we standardize and scale them to match SBERT's distribution of embedding values.
Refer to Appendix \autoref{app:training} for more information on training.

\begin{table*}[h!]
\centering
\begin{adjustbox}{max width=\textwidth}
    \begin{tabular}{llcccccccccccccccccccc}
        \toprule
        \multirow{3}{*}{\textbf{Dataset}} & \multirow{3}{*}{\textbf{Model}} & \multicolumn{10}{c}{\textbf{Traits}} & \multicolumn{4}{c}{\textbf{States}} & \multicolumn{6}{c}{\textbf{Dispositions}}\\
        
        \cmidrule(r){3-12} \cmidrule(r){13-16} \cmidrule(r){17-22}
        
        & & \multicolumn{2}{c}{\textbf{OPE}} & \multicolumn{2}{c}{\textbf{CON}} & \multicolumn{2}{c}{\textbf{EXT}} & \multicolumn{2}{c}{\textbf{AGR}} & \multicolumn{2}{c}{\textbf{NEU}} & \multicolumn{2}{c}{\textbf{VAL}} & \multicolumn{2}{c}{\textbf{ARO}} & \multicolumn{2}{c}{\textbf{ANG}} & \multicolumn{2}{c}{\textbf{ANX}} & \multicolumn{2}{c}{\textbf{DEP}}\\
        
        \cmidrule(r){3-4} \cmidrule(r){5-6} \cmidrule(r){7-8} \cmidrule(r){9-10} \cmidrule(r){11-12} \cmidrule(r){13-14} \cmidrule(r){15-16} \cmidrule(r){17-18} \cmidrule(r){19-20} \cmidrule(r){21-22}

        & & $r(\uparrow)$ & $mse(\downarrow)$ & $r$ & $mse$ & $r$ & $mse$ & $r$ & $mse$ & $r$ & $mse$ & $r$ & $mse$ & $r$ & $mse$ & $r$ & $mse$ & $r$ & $mse$ & $r$ & $mse$\\
        
        \midrule
        
        & \textit{SBERT-384} & .73 & \textbf{.11} & \textbf{.83} & \textbf{.07} & .68 & .11 & .75 & .06 & .77 & .09 & .69 & .001 & .81 & \textbf{.000} & .59 & \textbf{.03} & .60 & \textbf{.01} & .61 & .04\\

        \cmidrule{2-22}


        \multirow{7}{*}{HiTOP} & W2V2B & .63 & .14 & .69 & .12 & .75 & .09 & .60 & .09 & .72 & .10 & .65 & .001 & .73 & \textbf{.000} & .45 & .04 & .51 & .02 & .64 & \textbf{.03}\\

        & HuBERT & .67 & .13 & .71 & .11 & \textbf{.77} & \textbf{.08} & .57 & .10 & .70 & .11 & .66 & .001 & .73 & \textbf{.000} & .48 & .04 & .48 & .02 & .58 & .04\\
        
        & Whisper-384 & \textbf{.74} & \textbf{.11} & .80 & .08 & .69 & .10 & .76 & .06 & .78 & .08 & .71 & .001 & .82 & \textbf{.000} & .53 & \textbf{.03} & .61 & .01 & .65 & \textbf{.03}\\

        \cmidrule{2-22}
        
        & WhiSA-384 & .71$*$ & \textbf{.11} & .81$*$ & .08 & .70 & .10 & .77$*$ & .06 & .78$*$ & .08 & .73$*$ & .001 & .83$*$ & \textbf{.000} & .59$\dagger$ & \textbf{.03} & .61 & \textbf{.01} & .61 & .04\\
        
        & WhiSPA-384$_r$ & \textbf{.74$*$} & \textbf{.11} & \textbf{.83$\dagger$} & \textbf{.07} & .70 & .10 & \textbf{.79$\dagger$} & \textbf{.05} & .79$\dagger$ & \textbf{.07} & \textbf{.78$\dagger$} & \textbf{.000} & \textbf{.85$\dagger$} & \textbf{.000} & .59$\dagger$ & \textbf{.03} & .61$\dagger$ & \textbf{.01} & \textbf{.66$*$} & \textbf{.03}\\
        
        & WhiSPA-394 & .72$*$ & \textbf{.11} & \textbf{.83$\dagger$} & \textbf{.07} & .72 & .09 & \textbf{.79$\dagger$} & \textbf{.05} & \textbf{.82$\dagger$} & \textbf{.07} & .76$\dagger$ & \textbf{.000} & .84$*$ & \textbf{.000} & \textbf{.62$\dagger$} & \textbf{.03} & \textbf{.65$\dagger$} & \textbf{.01} & .63$*$ & \textbf{.03}\\
        
        \midrule[1.5pt]

        & \textit{SBERT-384} & .65 & .35 & .78 & .29 & .73 & .33 & .73 & .25 & .73 & .34 & .62 & .003 & \textbf{.86} & \textbf{.002} & .62 & .14 & .56 & .11 & .59 & .11\\

        \cmidrule{2-22}


        \multirow{7}{*}{WTC} & W2V2B & .33 & .54 & .51 & .55 & .34 & .64 & .37 & .46 & .34 & .64 & .32 & .004 & .51 & .005 & .31 & .21 & .14 & .16 & .22 & .15\\

        & HuBERT & .35 & .54 & .57 & .50 & .39 & .61 & .44 & .43 & .42 & .60 & .38 & .003 & .53 & .005 & .36 & .20 & .15 & .16 & .22 & .16\\
        
        & Whisper-384 & .57 & .43 & .70 & .37 & .68 & .38 & .64 & 32 & .67 & .40 & .56 & .003 & .82 & \textbf{.002} & .54 & .16 & .46 & .13 & .45 & .13\\

        \cmidrule{2-22}

        & WhiSA-384 & .70$\dagger$ & .31 & .82$\dagger$ & .24 & .75$\dagger$ & .32 & .76$\dagger$ & .23 & .77$\dagger$ & .30 & .67$\dagger$ & \textbf{.002} & .85$\dagger$ & \textbf{.002} & .66$\dagger$ & .13 & .61$\dagger$ & .10 & .61$\dagger$ & .10\\

        & WhiSPA-384$_r$ & .71$\dagger$ & .29 & .82$\dagger$ & .24 & .74$\dagger$ & .30 & .76$\dagger$ & .20 & .76$\dagger$ & .27 & .68$\dagger$ & \textbf{.002} & .85$\dagger$ & \textbf{.002} & .67$\dagger$ & .01 & .61$\dagger$ & \textbf{.09} & .61$\dagger$ & \textbf{.09}\\

        & WhiSPA-394 & \textbf{.72$\dagger$} & \textbf{.28} & \textbf{.83$\dagger$} & \textbf{.22} & \textbf{.76$\dagger$} & \textbf{.29} & \textbf{.79$\dagger$} & \textbf{.19} & \textbf{.79$\dagger$} & \textbf{.26} & \textbf{.70$\dagger$} & \textbf{.002} & \textbf{.86$\dagger$} & \textbf{.002} & \textbf{.69$\dagger$} & \textbf{.11} & \textbf{.64$\dagger$} & \textbf{.09} & \textbf{.66$\dagger$} & \textbf{.09}\\
        
        \bottomrule
    \end{tabular}
\end{adjustbox}
\caption{\textbf{Self-Supervised Prediction Accuracies for Psychological Traits, States, and Dispositions.} Averaged person-level embeddings were fit to a ridge regression with 10-fold cross validation. \textbf{Bold} indicates the best metric for the psychological scale in the respective dataset. $\uparrow$ implies \textit{higher} is \textit{better}. $\downarrow$ implies \textit{lower} is \textit{better}. $*$ indicates \textit{statistically significant ($p < .05$)} predictions compared to W2V2B. $\dagger$ indicates \textit{statistically significant ($p < .05$)} predictions compared to Whisper-384.
\label{tab:selfsupervised_eval}}
\end{table*}

\section{Results \& Discussion}
We consider three popular, robust speech encoders as baselines: Wav2Vec2-BERT\footnote{W2V2B version: \href{https://huggingface.co/hf-audio/wav2vec2-bert-CV16-en}{\texttt{wav2vec2-bert-CV16-en}}} ~\cite{communication2023w2v2bert, chung2021w2vbert}, HuBERT\footnote{HuBERT version: \href{https://huggingface.co/facebook/hubert-large-ls960-ft}{\texttt{hubert-large-ls960-ft}}} \cite{hsu2021hubert}, and Whisper \cite{radford2022whisper}, which are referred to as \textbf{W2V2B}, \textbf{HuBERT}, and \textbf{Whisper-384}, respectively.
We measured the effectiveness of these embeddings by computing Pearson correlation coefficient ($r$) and mean squared error ($mse$) over a 10-fold cross-validated ridge regression model for each task.
For each model variant, we encode audio segments for each participant and aggregate them with a statistical mean to represent person-level embeddings for the tasks in  \autoref{tab:selfsupervised_eval} and \autoref{tab:downstream_audio}. 

\paragraph{Alignment improved the models' ability to capture psychological dimensions from language.}
We evaluated the speech-based models' ability to capture the psychological dimensions of language by comparing our models' predictions to \psychemb derived values at the segment level. 
As summarized in \autoref{tab:selfsupervised_eval}, we found that both semantic (WhiSA) and psychological alignments (WhiSPA) significantly outperformed traditional speech-based models (W2V2B and Whisper) across all ten dimensions on both metrics.
Compared to Whisper, which was evidently a stronger baseline than W2V2B ($Avg\space\Delta\space = 36$ Pearson points for WTC \& $21$ points for HiTOP), 
Our semantic alignment method showed a marked improvement in performance, with an average of $11$ in Pearson points for WTC and $2$ in HiTOP.
A paired t-test was used to confirm that all improvements over W2V2B and all improvements over Whisper, except for 4 outcomes in HiTOP, were statistically significant ($p<.05$).   
This result highlighted our alignment methods improved the speech model's ability to capture psychological dimensions in language (\psychemb). 

Interestingly, deriving psychological estimates from semantic dimensions (WhiSPA-394) was consistently better than the replacement (WhiSPA-384$_r$) of 10 semantic dimensions with \psychemb dimension.
This shows the importance of curating the semantic dimensions before replacing them with different embeddings.  

\begin{figure}[h!]
    \centering
    \begin{subfigure}[b]{.25\textwidth}
        \centering
        \includegraphics[width=\textwidth]{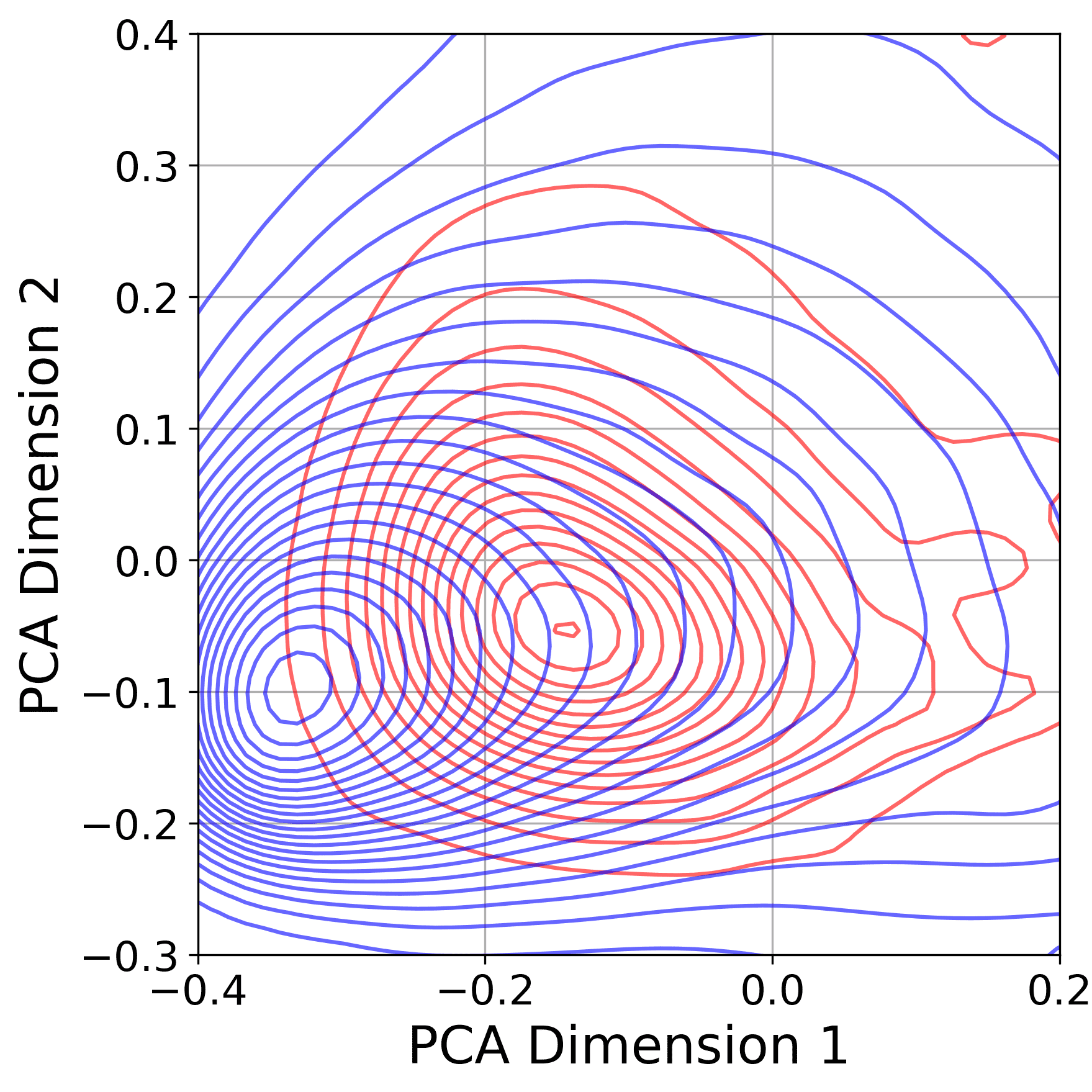}
        \caption{Before Alignment}
        \label{fig:alignment_comparison_before}
    \end{subfigure}%
    \begin{subfigure}[b]{.25\textwidth}
        \centering
        \includegraphics[width=\textwidth]{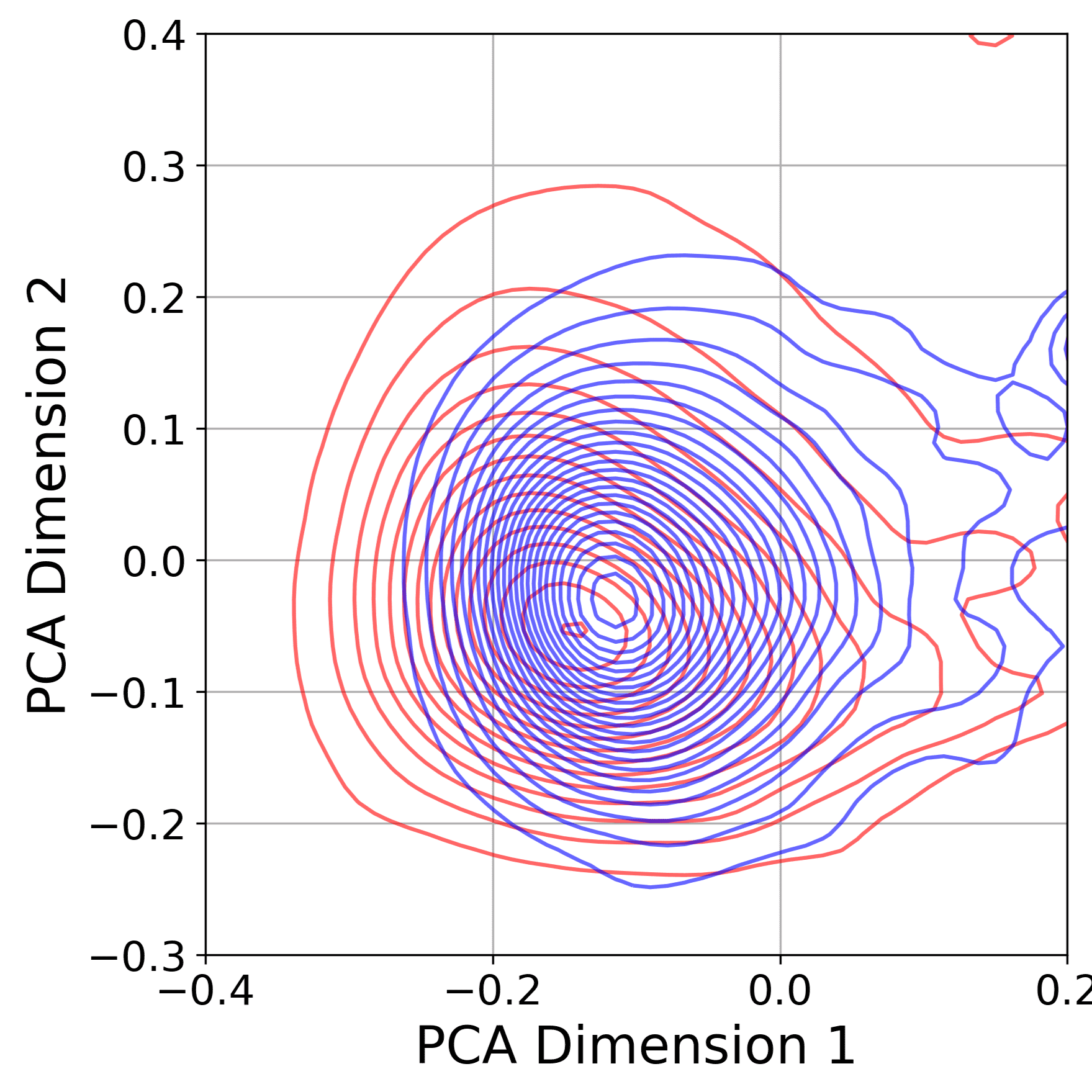}
        \caption{After Alignment}
        \label{fig:alignment_comparison_after}
    \end{subfigure}
    \caption{Bivariate KDE contour plot of PCA dimensionally reduced speech/text embeddings. Speech representations in \textcolor{blue}{blue}. Text representations in \textcolor{red}{red}.}
    \label{fig:alignment_comparison}
\end{figure}

We also observed that the alignment increased the overlap between the latent space of the speech and text embeddings, as shown in~\autoref{fig:alignment_comparison}. 
Before alignment (\autoref{fig:alignment_comparison_before}), speech and text embeddings show distinct contours with very little overlap in their dense regions, highlighting a clear modality gap and a lack of shared contextual meaning. 
After alignment (\autoref{fig:alignment_comparison_after}), the contours exhibit greater overlap, indicating a unified embedding space with reduced variance.
\autoref{fig:alignment_comparison} demonstrates that the alignment process effectively bridges the semantic gap between the two modalities.

\begin{table*}[h!]
\centering
\begin{adjustbox}{max width=\textwidth}
    \begin{tabular}{lcccccccccccccccccccccc}
        \toprule
        \multirow{3}{*}{\textbf{Model}} & \multicolumn{10}{c}{\textbf{HiTOP}} & \multicolumn{12}{c}{\textbf{WTC}} \\
        
        \cmidrule(r){2-13} \cmidrule(r){14-23}
        
        & \multicolumn{2}{c}{\textbf{INT}} & \multicolumn{2}{c}{\textbf{DIS}} & \multicolumn{2}{c}{\textbf{ANT}} & \multicolumn{2}{c}{\textbf{SOM}} & \multicolumn{2}{c}{\textbf{THD}} & \multicolumn{2}{c}{\textbf{DET}} & \multicolumn{2}{c}{\textbf{PCL}} & \multicolumn{2}{c}{\textbf{REX}} & \multicolumn{2}{c}{\textbf{AVO}} & \multicolumn{2}{c}{\textbf{NAM}} & \multicolumn{2}{c}{\textbf{HYP}}\\

        \cmidrule(r){2-3} \cmidrule(r){4-5} \cmidrule(r){6-7} \cmidrule(r){8-9} \cmidrule(r){10-11} \cmidrule(r){12-13} \cmidrule(r){14-15} \cmidrule(r){16-17} \cmidrule(r){18-19} \cmidrule(r){20-21} \cmidrule(r){22-23}

        & $r(\uparrow)$ & $mse(\downarrow)$ & $r$ & $mse$ & $r$ & $mse$ & $r$ & $mse$ & $r$ & $mse$ & $r$ & $mse$ & $r$ & $mse$ & $r$ & $mse$ & $r$ & $mse$ & $r$ & $mse$ & $r$ & $mse$\\
        
        \midrule
        
        W2V2B & .50 & .17 & .46 & .21 & .35 & .11 & -.00 & .24 & .27 & .11 & .32 & .20 & .14 & 133.19 & .14 & 12.08 & .07 & 3.99 & .13 & 10.98 & .10 & 17.17\\

        HuBERT & .50 & .17 & .53 & .19 & .36 & .11 & .07 & .23 & .28 & .11 & .31 & .20 & .21 & 129.86 & .22 & 11.72 & .07 & 3.99 & .19 & 10.80 & .15 & 16.99\\
        
        Whisper-384 & .39 & .19 & .33 & .24 & .33 & .11 & .07 & .23 & .28 & .11 & .29 & .20 & .23 & 128.85 & .21 & 11.77 & .06 & 4.00 & .19 & 10.87 & .23 & 16.41\\

        \midrule

        WhiSA-384 & .55$\dagger$ & .16 & .53$\dagger$ & \textbf{.19} & \textbf{.43$\dagger$} & \textbf{.10} & .22$\dagger$ & .23 & .37$\dagger$ & \textbf{.10} & .33$\dagger$ & \textbf{.18} & .29$\dagger$ & 119.68 & .27$\dagger$ & 11.26 & .19$\dagger$ & 3.90 & .26$\dagger$ & 10.12 & .28$\dagger$ & 15.56\\

        WhiSPA-384$_r$ & .56$\dagger$ & \textbf{.15} & .53$\dagger$ & \textbf{.19} & .42$\dagger$ & \textbf{.10} & \textbf{.23$*$} & \textbf{.22} & \textbf{.39$\dagger$} & \textbf{.10} & \textbf{.39$\dagger$} & .19 & .34$\dagger$ & 119.24 & \textbf{.30$\dagger$} & 11.23 & .17 & 3.88 & .31$\dagger$ & \textbf{10.08} & \textbf{.32$\dagger$} & 15.54\\
        
        WhiSPA-394 & \textbf{.57$\dagger$} & \textbf{.15} & \textbf{.54$\dagger$} & \textbf{.19} & \textbf{.43$\dagger$} & \textbf{.10} & .22$\dagger$ & \textbf{.22} & .37$\dagger$ & \textbf{.10} & .38$\dagger$ & .19 & \textbf{.35$\dagger$} & \textbf{118.91} & \textbf{.30$\dagger$} & \textbf{11.18} & \textbf{.20} & \textbf{3.85} & \textbf{.32$\dagger$} & 10.09 & \textbf{.32$\dagger$} & \textbf{15.48}\\
        \bottomrule
    \end{tabular}
\end{adjustbox}
\caption{\textbf{Self-Reported/Annotated Prediction Accuracies for Psychological Scales.} Averaged person-level embeddings were fit to a ridge regression with 10-fold cross validation. \textbf{Bold} indicates the best metric for the psychological scale in the respective dataset. $\uparrow$ implies \textit{higher} is \textit{better}. $\downarrow$ implies \textit{lower} is \textit{better}. $*$ indicates \textit{statistically significant ($p < .05$)} predictions compared to W2V2B. $\dagger$ indicates \textit{statistically significant ($p < .05$)} predictions compared to Whisper-384.
\label{tab:downstream_audio}}
\end{table*}

\begin{table*}[hb!]
\centering
\small
\begin{adjustbox}{center=\textwidth}
    \begin{tabular}{lcccccc}
        \toprule
        \multirow{2}{*}{\textbf{Model}} & \multirow{2}{*}{\textbf{Loss}} & \multicolumn{2}{c}{\textbf{Self-Supervision Tasks}} & \multicolumn{2}{c}{\textbf{Downstream Tasks}}\\ 
        \cmidrule(r){3-4} 
        \cmidrule(r){5-6} 
        & & Pearson $r$ ($\uparrow$) & MSE ($\downarrow$) & Pearson $r$ ($\uparrow$) & MSE ($\downarrow$)\\ 
        \midrule
        \multirow{2}{*}{WhiSA-384} & \textit{CS} & .72 & .11 & .34 & 15.26\\
        & \textit{NCE} & .72 & .11 & .36 & 14.63\\
        \midrule
        WhiSPA-384$_r$ & \textit{CS} & .72 & .12 & .34 & 15.08\\
        \textit{(with replacement)} & \textit{NCE} & .73 & .11 & .36 & 14.68\\
        \midrule
        WhiSPA-394 & \textit{CS} & .72 & .11 & .34 & 15.21\\
        \textit{(with projection)} & \textit{NCE} & \textbf{.74} & \textbf{.10} & \textbf{.37} & \textbf{14.59}\\
        \bottomrule
    \end{tabular}
\end{adjustbox}
\caption{\textbf{Comparison of Loss Functions on Self-Supervised and Downstream Tasks.} The reported Pearson $r$'s and MSE's are averaged across all outcomes. \textbf{Bold} indicates the best metric when comparing loss functions across different models. $\uparrow$ implies \textit{higher} is \textit{better}. $\downarrow$ implies \textit{lower} is \textit{better}.
\label{tab:loss_comparison}}
\end{table*}

\paragraph{Semantic-Psychological alignment is SotA for speech-based psychological assessments.}
\autoref{tab:downstream_audio} shows that the improvements brought by our aligned models over traditional models were preserved even when evaluated on a spectrum of downstream psychological assessment tasks.
In particular, the alignment showed a stark increase in capturing deeper psychological conditions such as \textbf{INT} (internalizing) ($\geq 16$ Pearson points) and \textbf{DIS} (disinhibition) ($\geq 20$ Pearson points) from very long durations of speech data. 
Consistent with behaviours exhibited with \psychemb dimensions, in \autoref{tab:selfsupervised_eval}, semantic-psychological alignment from semantically-derived psychological dimensions (WhiSPA-394) performed the best, followed by semantic-psychological alignment from replacement (WhiSPA-384$_r$) and finally semantic-only alignment (WhiSA-384). 
For these tasks, we averaged the segment-level representations of the interview audio file to produce a person-level embedding.  
These embeddings were used to perform 10-fold cross-validation with a ridge regression model, and its performance was measured using Pearson correlation coefficient ($r$) and mean squared error ($mse$).

The success of WhiSPA-394 can be attributed to its integration of psychological feature alignment, which complements semantic alignment by explicitly encoding affective dimensions such as valence and arousal. 
The improvements in outcomes like \textbf{INT} and \textbf{DIS} further support this interpretation since these constructs often rely on subtle vocal cues, such as pause distribution, pitch variability, and vocal tone as established by prior works \cite{kotov2024-hitop}.
By injecting dimensions with psychological relevance into the alignment process, the model bridges the gap between the prosodic information in speech and the textual semantics used to train baseline models like WhiSA.
This dual alignment likely enhances the model’s ability to capture both the what (semantic content) and the how (affective delivery) of speech, enabling more accurate predictions of psychological scales.

\begin{table*}[ht]
\centering
\begin{adjustbox}{max width=\textwidth}
    \begin{tabular}{lcccccccccccccccccccccc}
        \toprule
        \multirow{3}{*}{\textbf{Model}} & \multicolumn{10}{c}{\textbf{HiTOP}} & \multicolumn{12}{c}{\textbf{WTC}} \\
        
        \cmidrule(r){2-13} \cmidrule(r){14-23}
        
        & \multicolumn{2}{c}{\textbf{INT}} & \multicolumn{2}{c}{\textbf{DIS}} & \multicolumn{2}{c}{\textbf{ANT}} & \multicolumn{2}{c}{\textbf{SOM}} & \multicolumn{2}{c}{\textbf{THD}} & \multicolumn{2}{c}{\textbf{DET}} & \multicolumn{2}{c}{\textbf{PCL}} & \multicolumn{2}{c}{\textbf{REX}} & \multicolumn{2}{c}{\textbf{AVO}} & \multicolumn{2}{c}{\textbf{NAM}} & \multicolumn{2}{c}{\textbf{HYP}}\\

        \cmidrule(r){2-3} \cmidrule(r){4-5} \cmidrule(r){6-7} \cmidrule(r){8-9} \cmidrule(r){10-11} \cmidrule(r){12-13} \cmidrule(r){14-15} \cmidrule(r){16-17} \cmidrule(r){18-19} \cmidrule(r){20-21} \cmidrule(r){22-23}

        & $r(\uparrow)$ & $mse(\downarrow)$ & $r$ & $mse$ & $r$ & $mse$ & $r$ & $mse$ & $r$ & $mse$ & $r$ & $mse$ & $r$ & $mse$ & $r$ & $mse$ & $r$ & $mse$ & $r$ & $mse$ & $r$ & $mse$\\
        
        \midrule
        \multicolumn{6}{l}{\textbf{Transcribe $\rightarrow$ Text Model} (\textit{Upperbound})}\\
        
        \textit{SBERT-384} & .54* & .16 & \textbf{.55*} & \textbf{.19} & \textbf{.43*} & \textbf{.10} & .16* & .23 & \textbf{.40*} & \textbf{.10} & \textbf{.40*} & \textbf{.18} & \textbf{.36*} & \textbf{118.21} & \textbf{.32} & \textbf{11.08} & \textbf{.24*} & \textbf{3.80} & \textbf{.32*} & \textbf{10.08} & \textbf{.33*} & \textbf{15.43}\\

        \textit{SBERT-1024} & .65* & \textbf{.13} & .59* & .17 & .51* & \textbf{.09} & .20 & .23 & \textbf{.43*} & \textbf{.09} & \textbf{.44*} & \textbf{.18} & .37* & 118.67 & .32* & 10.99 & \textbf{.27*} & \textbf{3.68} & .31* & 10.26 & .31* & 15.66\\
        
        \midrule
        \textbf{Audio Model}\hspace*{4em}\\
        Whisper-384 & .39 & .19 & .33 & .24 & .33 & .11 & .07 & .23 & .28 & .11 & .29 & .20 & .23 & 128.85 & .21 & 11.77 & .06 & 4.00 & .19 & 10.87 & .23 & 16.41\\

        WhiSPA-394 & \textbf{.57*} & \textbf{.15} & .54* & \textbf{.19} & \textbf{.43*} & \textbf{.10} & \textbf{.22*} & \textbf{.22} & .37* & \textbf{.10} & .38* & .19 & .35* & 118.91 & .30* & 11.18 & .20 & 3.85 & \textbf{.32*} & 10.09 & .32* & 15.48\\

        \midrule
        \multicolumn{6}{l}{\textbf{Audio Model (Scaled Up)}}\\
        Whisper-1024 & .57 & .15 & .52 & .19 & .44 & .10 & \textbf{.23} & \textbf{.22} & .34 & .10 & .37 & .19 & .28 & 126.03 & .23 & 11.81 & .11 & 3.96 & .27 & 10.54 & .26 & 16.29\\
        
        WhiSPA-1034 & \textbf{.67*} & \textbf{.13} & \textbf{.62*} & \textbf{.16} & \textbf{.53*} & .10 & .21 & \textbf{.22} & .40* & .10 & \textbf{.44*} & \textbf{.18} & \textbf{.41*} & \textbf{114.44} & \textbf{.34*} & \textbf{10.89} & \textbf{.27*} & 3.70 & \textbf{.37*} & \textbf{9.84} & \textbf{.34*} & \textbf{15.42}\\
        
        \bottomrule
    \end{tabular}
\end{adjustbox}
\caption{\textbf{Performance of WhiSPA Distilled Across Larger Dimensionalities.} Averaged person-level embeddings were fit to a ridge regression with 10-fold cross validation. \textbf{Bold} indicates the best metric for the psychological scale in the respective dataset. $\uparrow$ implies \textit{higher} is \textit{better}. $\downarrow$ implies \textit{lower} is \textit{better}. $*$ indicates \textit{statistically significant ($p < .05$)} predictions compared to Whisper-(384/1024).
\label{tab:whispa_scaling}}
\end{table*}

\paragraph{Contrastive loss criteria led to richer representations of audio.}
Investigation of the choice of alignment objective towards performance (\autoref{tab:loss_comparison}) revealed that Noise Contrastive Estimation (NCE) consistently produced a better-aligned model than cosine similarity (CS).
This is likely because NCE optimizes for discriminative learning, encouraging more separation between positive and negative samples in the embedding space~\cite{ye2022crossmodalspeech}, enhancing the model's ability to encode nuanced semantic and psychological cues. 
When comparing WhiSPA-394 and WhiSPA-384, we notice the recurring trend with NCE granting a greater optima during alignment than CS as exemplified in \autoref{tab:loss_comparison}.
However, WhiSPA-384 holds its ground in HiTOP, achieving comparable correlations.
This suggests that WhiSPA-394's architecture may generalize well to diverse datasets but thrives in highly semantic and affective audio contexts like WTC.

\paragraph{WhiSPA effectively scales to larger dimensionalities.}
To investigate the effects of utilizing a larger teacher LM, we conducted experiments with \texttt{all-roberta-large-v1} ($\sim$330M) paired with \texttt{whisper-medium} ($\sim$796M) as the audio backbone, each with 1024 embedding dimensions.
\autoref{tab:whispa_scaling} shows that the distillation process remains effective for aligning larger student-teacher model configurations, further validating its scalability and generalizability for the downstream task.
When comparing Whisper-384 to WhiSPA-394, we observe an \textbf{average error reduction of 83.38\%}, while the 1024-sized models show an \textbf{even greater reduction of 86.61\%.}
A larger audio backbone also improves the consistency with which the student model outperforms its language-based teacher, likely due to enhanced context retention afforded by a larger parameter space.
For example, WhiSPA-394 surpasses its teacher in only 2 of 11 outcomes, whereas WhiSPA-1034 does so in 7 of 11 psychological assessments.
These findings underscore the effectiveness of our distillation strategy, particularly for larger models offering greater embedding dimensionality.

\begin{table}[h!]
\centering
\small
\begin{tabular}{lcccccc}
    \toprule
    \textbf{Model} & \textbf{PCL} & \multicolumn{3}{c}{\textbf{HiTOP}} & \textbf{VAL}\\
    \cmidrule(r){3-5}
    & & INT & DIS & THD & \textit{(segment)} \\
    \midrule
    
    SBERT-384 & \textbf{.36} & .54 & .55 & \textbf{.40} & .47\\
    
    \cmidrule(r){1-6}
    
    Whisper-384 & .23 & .39 & .33 & .28 & .38\\

    \cmidrule(r){1-6}
    
    WhiSA-384 & .29 & .55 & .53 & .37 & .50*\\
    
    WhiSPA-384$_r$ & .34 & .56* & .53 & .39 & \textbf{.53*}\\
    
    WhiSPA-394 & .35 & .57* & .54 & .37 & .51*\\
    
    \cmidrule(r){1-6}
    
    WhiSPA-394 & \multirow{2}{*}{\textbf{.36}} & \multirow{2}{*}{\textbf{.58*}} & \multirow{2}{*}{\textbf{.56}} & \multirow{2}{*}{.39} & \multirow{2}{*}{.52*}\\
    \& SBERT-384\\
    
    \bottomrule
\end{tabular}
\caption{\textbf{Comparison of Audio and Text Models for Predicting Psychological Scales.} Acoustic valence (\textbf{VAL}) was regressed on 300 human-annotated audio segments. SBERT-384 utilizes a cascaded pipeline (Whisper transcript $\rightarrow$ SBERT encoding). \textit{Higher} is \textit{Better}. * indicates statistically significant ($p < .05$) predictions compared to SBERT-384.
\label{tab:all_segment_eval}}
\end{table}

\begin{figure*}[t]
\centering
\includegraphics[width=\textwidth]{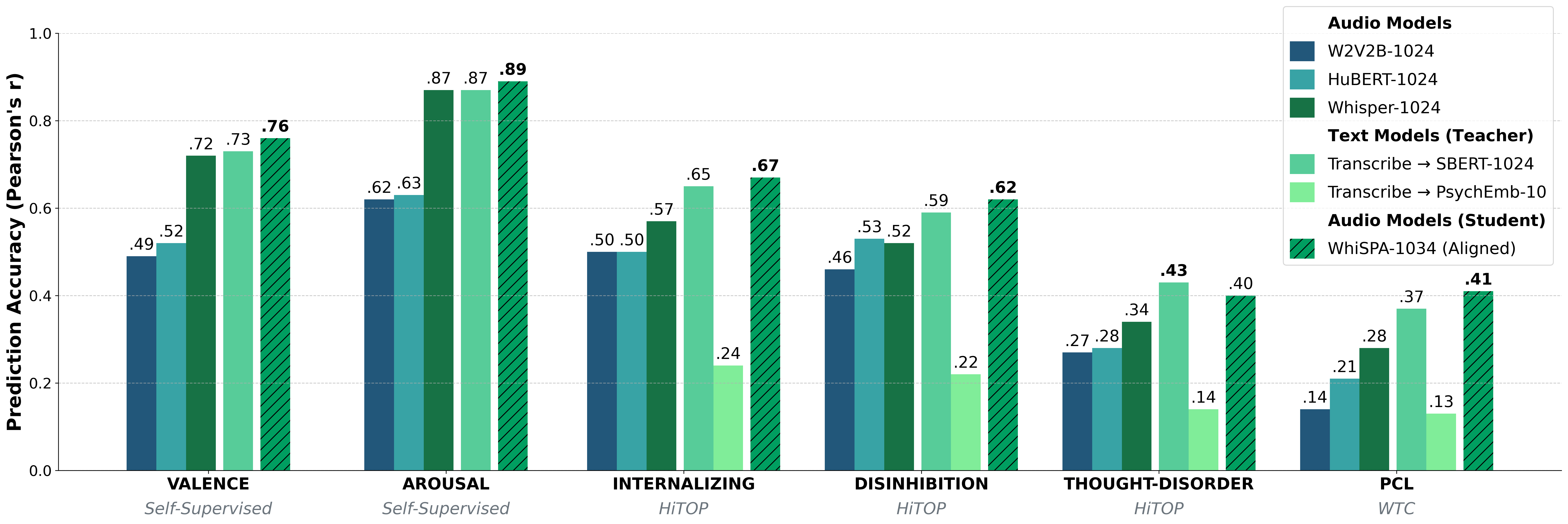}
\caption{\textbf{WhiSPA Bridges the Semantic/Psychological Representation Gap.} WhiSPA consistently outperforms every baseline audio model and, in most cases, matches or exceeds the performance of the text-based language model teacher.}
\label{fig:model_bars}
\end{figure*}

\paragraph{WhiSPA captures semantics without the need for appending SBERT representations.}
The last row in \autoref{tab:all_segment_eval} underscores the marginal increase in correlations after appending SBERT embeddings to WhiSPA.
WhiSPA, trained through a student-teacher alignment paradigm, appears to reach a semantic and psychological optimum during convergence.
This is evident in its substantial performance gains over Whisper, which lacks the semantic and psychological depth provided by language models.
However, the potential of cross-modal alignment may be constrained by the representational efficacy of the teacher model(s).
On human-annotated audio segments, all of the WhiSPA variants achieve substantial improvements in capturing acoustic valence.
In comparison with Whisper-384, WhiSPA-384$_r$ exhibits a gain of $+15$ Pearson points in \textbf{VAL} (acoustic valence) which exemplifies the reduction in the semantic/psychological gap between audio models and text-based models. 
Notably in \autoref{fig:model_bars}, WhiSPA-1034 demonstrates clear improvements in the majority of outcomes, with an average gain of $+2$ Pearson points, when compared to its teacher, SBERT-1024.

Ultimately, these findings highlight \textbf{two important observations:}
(1) WhiSPA effectively captures nearly all the information encoded by its text-based teacher model, SBERT.
(2) The marginal returns from appending text-based representations indicate that WhiSPA successfully learns to encode the critical semantic and psychological cues provided by its teachers, reflecting the success of the distillation.

\paragraph{WhiSPA's representations are interpretable through language semantically associated with psychological dimensions.}
\autoref{tab_app:quant_ngrams_pos} shows that n-grams known to be indicative of PTSD severity from prior studies~\cite{kjell2024wtc} —- including first-person pronouns, experienced symptoms, psychological distress, and negative affect -- yield significantly higher correlations with WhiSPA’s predictions compared to Whisper. In contrast, \autoref{tab_app:quant_ngrams_neg} reveals that language discussing relationships and positive affect is more negatively associated with WhiSPA’s scores. 
These findings indicate that the contrastive loss training effectively aligns the latent space with rich semantic and psychological representations, capturing psychologically relevant linguistic markers more robustly.
The highly semantic latent spaces of text-based LMs are reflected in WhiSPA's representations, especially for psychological nuances in spoken language.
More quantitative analysis of our model can be found in Appendix \autoref{app:quant_analysis}

\section{Conclusion}
We claim that WhiSPA is a significant step toward more accurate representations of human communication by addressing the modal gap between text and audio, as language models often outperform audio models in predicting psychological attributes.
By aligning WhiSPA’s representations with SBERT’s representations enriched with \psychemb, we found consistent improvement for ten self-supervised tasks and significantly greater accuracies over 11 downstream psychological tasks.
We observed only marginal improvements when appending SBERT representations to WhiSPA's, implying that the distillation process effectively captures the semantic features provided by the teacher language model.
Our findings exemplify WhiSPA's effectiveness in extracting semantic and psychological features from speech, enhancing SotA audio representations for psychological and mental health assessments.

\section{Limitations}
While WhiSPA demonstrates significant advancements in providing semantically enriched audio embeddings, its current training paradigm predominantly aligns with psychological features derived from text, potentially limiting its capacity to capture critical acoustic information.
This lexical bias, while beneficial for aligning with language-based models, raises an important question: \textit{to what extent can WhiSPA’s embeddings be further refined to incorporate affective context for psychological prediction?}
Given that vocal prosody and acoustic features convey essential emotional and psychological cues beyond textual content \cite{low2020-diagnosis}, incorporating these dimensions is crucial for a more comprehensive representation.

We acknowledge that this strong alignment with text-based language models may introduce an imbalance, diminishing the richness of acoustic cues that are particularly valuable for affective and psychological assessments.
Despite WhiSPA’s demonstrated success-—matching its language model teacher in psychological prediction and surpassing state-of-the-art audio models—-there remains an opportunity to enhance its representational capacity by preserving acoustic features.
To address this, future work will explore a multi-weighted dual loss objective, ensuring that WhiSPA retains a broader spectrum of information beyond textual representations.
We suspect this refinement would not only improve its efficacy in psychological modeling but also enhance its versatility for general-purpose speech tasks like automatic speech recognition (ASR) and emotion recognition in conversation (ERC), where both linguistic and acoustic cues are essential.

\section{Ethical Implications}
The multimodal WhiSPA model holds significant potential for improving mental healthcare assessments by providing rich insights into individuals’ states of mind through speech analysis. However, multimodal approaches increase ethical considerations due to the richer and more diverse forms of personally identifiable information (PII) they capture compared to unimodal models. In addition to text content, the WhiSPA model processes acoustic and prosodic features — including tone of voice, speech patterns, and emotional expressions — which can inadvertently reveal sensitive details like gender, ethnicity, emotional state, and health conditions. This expanded data scope raises the risk of re-identification, making it essential to implement stringent data security and handling, including compliance with privacy regulations such as GDPR and HIPAA.

\paragraph{Security \& Privacy.} Moreover, the potential for misuse or unauthorized exploitation of such detailed multimodal data necessitates robust ethical guidelines for its storage, processing, and application. Transparency in how these models are trained and used is critical to building trust among clinicians and patients. Finally, ongoing efforts to mitigate algorithmic biases and ensure fairness are important, as errors in multimodal assessments could disproportionately impact vulnerable populations or lead to incorrect diagnoses if not carefully managed.

This work was part of a study approved by SBU's Institutional Review Board (IRB \#1157153 World Trade Center Responder Language and Health Study) and (IRB \#2022-00391: iHiTOP).
The WTC and HiTOP recordings took place in a clinical setting at the Stony Brook WTC Health and Wellness Program where each participant gave consent and was fully informed about the study, that it was voluntary to take part, and that they had the right to withdraw at any time without giving a reason or that it would affect their treatment. After the interview, participants were debriefed (for more details about the WTC data collection, see \cite{kjell2024wtc}; for more details about the HiTOP data, see \cite{kotov2022-hitop, kotov2024-hitop}. The studies and data uses were approved by the Institutional Review Board at an undisclosed university for privacy reasons.

\paragraph{Software.} Adhering to the ideals of open and reproducible science, we will make the WhiSPA software code base, along with the trained models and secure dimensional representations of the data, openly available. These representations strictly comply with established security protocols, ensuring that no individual can be identified nor any anonymity safeguard compromised. Nevertheless, direct access to the underlying data remains restricted in accordance with privacy and security measures.

Additionally, AI-based tools were employed throughout the project to assist in code development and report formulation, including the use of ChatGPT and other similar consumer generative AI. Such integration aligns with established best practices and guidelines, ensuring that the technical accuracy, integrity, and scientific rigour of the work remain uncompromised while benefiting from enhanced efficiency and streamlined workflows.

\section*{Acknowledgments}
This work was supported in part by a grant from the CDC/NIOSH (U01 OH012476) and a grant from the NIH-NIAAA (R01 AA028032).
The conclusions contained herein are those of the authors and should not be interpreted as necessarily representing the official policies, either expressed or implied of the CDC, NIH, or any other government organization.
We greatly thank the participants, creators, maintainers, and interviewers from the WTC and HiTOP studies for enabling this research.

\newpage
\bibliography{references, anthology}

\appendix
\section{Appendix}
\label{sec:appendix}

\subsection{Data Description}
\label{app:data-description}

\subsubsection{HiTOP.}
The HiTOP dataset consists of video-recorded interviews conducted between World Trade Center responder participants and clinicians. Each recording is annotated with the outcomes derived from the HiTOP structured interview, which includes a standardized set of questions designed to assess a comprehensive set of mental health dimensions, including aspects of internalizing (e.g., questions about distress and fear), dis-inhibited externalizing (e.g., questions about substance abuse and antisocial behaviors) and more. This dataset comprises of 525 unique participants. 

\paragraph{Outcomes in HiTOP}
The HiTOP outcomes were derived from the structured clinical interview \cite{roman-meyer-2024-analysis}, where we used the total score of the six dimensions including:
i) internalizing (INT; e.g., dysphoria, lassitude),
ii) disinhibited externalizing (DIS; e.g., alcohol use, drug use), 
iii) antagonistic externalizing (ANT; e.g., attention seeking, callousness), 
iv) somatoform (SOM; e.g., conversion, somatization), 
v) thought disorder (THD; e.g., psychotic and disorganized thought patterns),
vi) detachment (DET; e.g., intimacy avoidance, suspiciousness)

\begin{figure}[ht]
\includegraphics[width=\columnwidth]{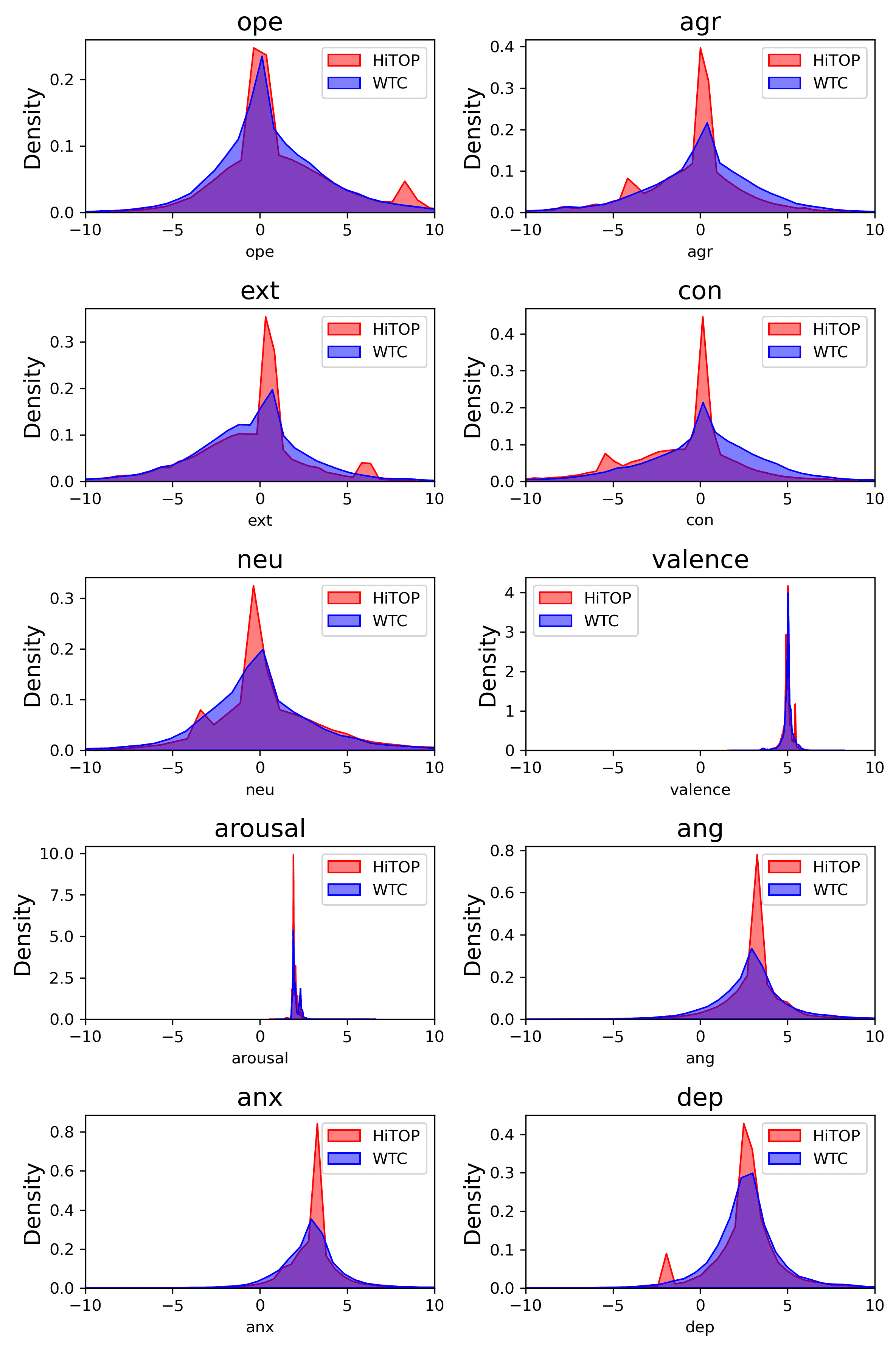}
\caption{Standardized distributions of \psychemb dimensions for each segment across both datasets. The distribution of WTC is shown in \textcolor{blue}{blue}. The distribution of WTC is shown in \textcolor{red}{red}.}
\label{fig:psychlex_dist}
\end{figure}

\subsubsection{WTC.}
In the WTC dataset, participants were recorded in a private room during their clinical visit while responding to questions displayed on a screen as part of an automated clinical interview. These questions prompted participants to reflect on both positive (e.g., What are three things you currently look forward to the most?) and negative aspects of their lives across different time frames (past, present, and future). Topics included general life experiences (e.g., the best and worst experiences, challenges, and support systems) and significant events such as COVID-19 and 9/11 (e.g., How does 9/11 affect you now?). A full list of the questions is provided in \cite{kjell2024wtc}.

To enhance generalizability, the questions were designed to be broad and used everyday language, avoiding clinical jargon or references to specific symptoms. Instructions on the screen advised participants not to read the questions aloud and to aim for at least 60 seconds of response time per question. Throughout the development phase, the questions were refined over three iterations to improve engagement and elicit more detailed responses. However, for the evaluation phase, the same set of questions was used for all participants. On average, recordings for those who met a threshold of at least 150 words lasted 7.5 minutes (SD = 4.1; range = 1.1 to 43.0 minutes). 

The data, from its source, totalled 1437 participants (Female = 7\%, Male = 93\%; Mean age = 57.9, SD = 8.0 years; 14.5\%).

\paragraph{Outcomes in WTC}
The PCL score and subscales were derived from the PTSD CheckList (PCL) \cite{blanchard1996-pcl}, which consists of 17 items designed to measure the severity of PTSD symptoms according to the Diagnostic and Statistical Manual of Mental Disorders, Fourth Edition (DSM-IV) criteria. Participants rate their experiences over the past month using a scale from 1 (not at all) to 5 (extremely). We calculated both the overall score (PCL) and scores for the four subscales. These subscales are Re-experiencing (REX; e.g., intrusive thoughts related to trauma), Avoidance (AVO; e.g., evading trauma-related thoughts), Emotional Numbing (NAM; e.g., difficulty recalling aspects of the trauma), and Hyperarousal (HYP; e.g., disturbances in sleep patterns). Reliability, as measured by Cronbach’s alpha, was acceptable across all scales ($\geq$ .70).

\subsection{Training}
\label{app:training}
The research done for devising WhiSPA's framework resulted from iterations of tweaking and testing architectures, loss criteria, parameters, and hyperparameters.

For the methodology presented in this paper, we provide the following configurations for reproducibility:\\
Pooling: $MEAN$. Learning Rate: $1 \times10^{-5}$. Weight Decay: $1 \times 10^{-2}$. Temperature ($\tau$): $0.1$. Batch Size: $900$. Number of Epochs: $50$. Number of workers (CPU cores): $16$. These configurations result in a total average training time of $\sim20$ hours.

We discovered that the efficacy of \autoref{nce-loss} highly depends on the batch size. It should be stated that larger batch sizes allow for greater degrees of repulsion and attraction in the cross-modal embedding space. While training WhiSA and WhiSPA, we utilized a batch size of $900$ and distributed them across $3$ NVIDIA RTX A6000 devices with 48GB of VRAM each.

Additionally, we use open-source licensed pre-trained models from \href{https://huggingface.co/}{HuggingFace}.
Our programmatic implementation for deep learning is done with \href{https://pytorch.org/}{PyTorch}.
When it comes to evaluation, we utilize \href{https://dlatk.github.io/dlatk/}{Differential Language Analysis Tool Kit (DLATK)} for correlating regression results across specified groups (i.e., $user\_id$ or $segment\_id$).

\begin{figure}[h!]
\includegraphics[width=\columnwidth]{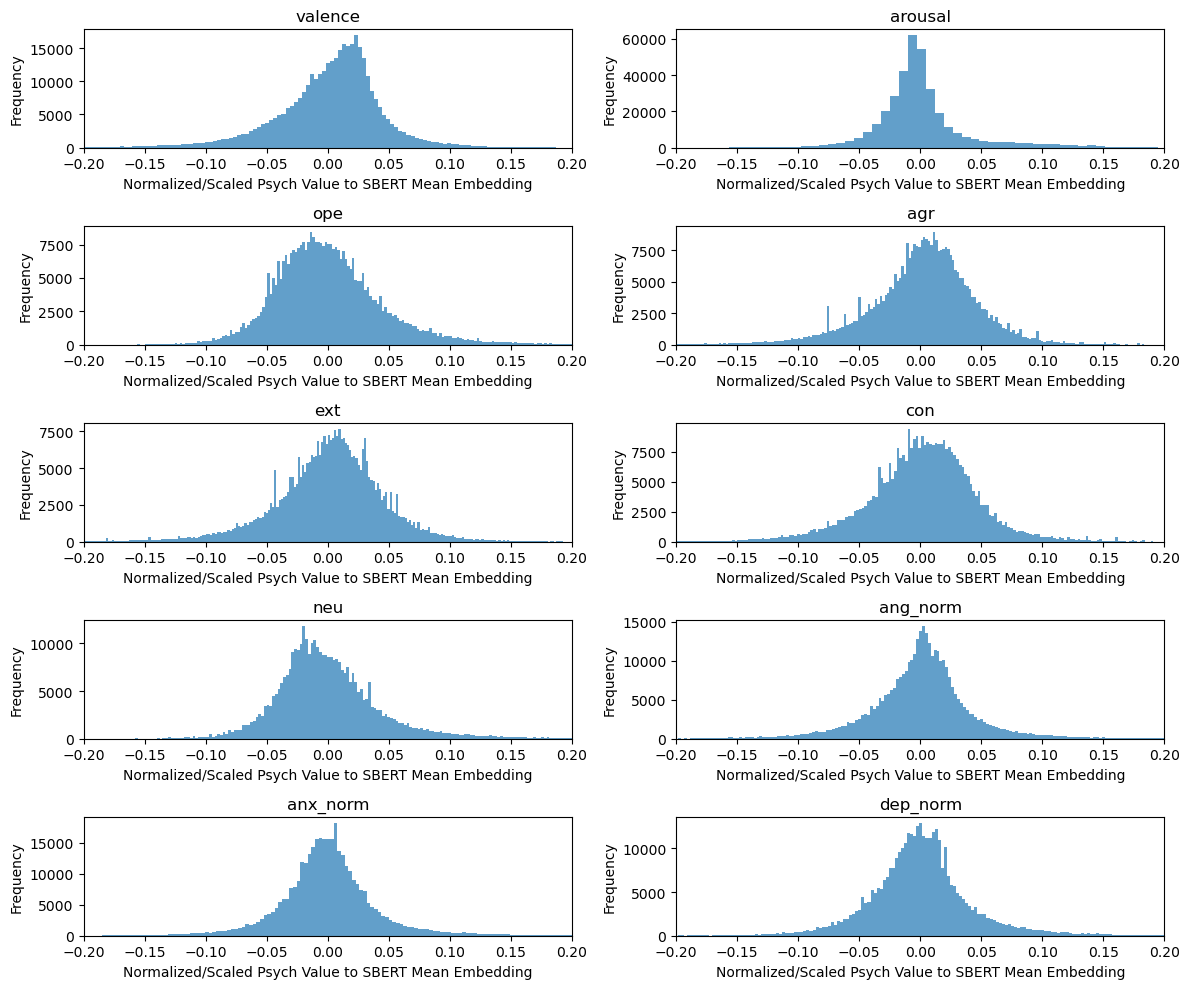}
\caption{Distributions of psychological features standardized and scaled to the distribution of SBERT's mean embedding value before augmentation for WhiSPA alignment training.}
\label{fig:psych_scaled_dist}
\end{figure}

Cosine similarity is sensitive to the relative magnitudes of the vectors being compared.
If the added ten dimensions of psychological features have a very different scale or distribution from SBERT embeddings as visualized in \autoref{fig:psych_scaled_dist}, they could dominate or skew the cosine similarity computation.
Once either loss function is applied, \eqref{cs-loss} or \eqref{nce-loss}, WhiSPA embeddings remain semantically aligned with SBERT while also encoding meaningful affective cues for downstream tasks.

\begin{figure}[h]
\includegraphics[width=\columnwidth]{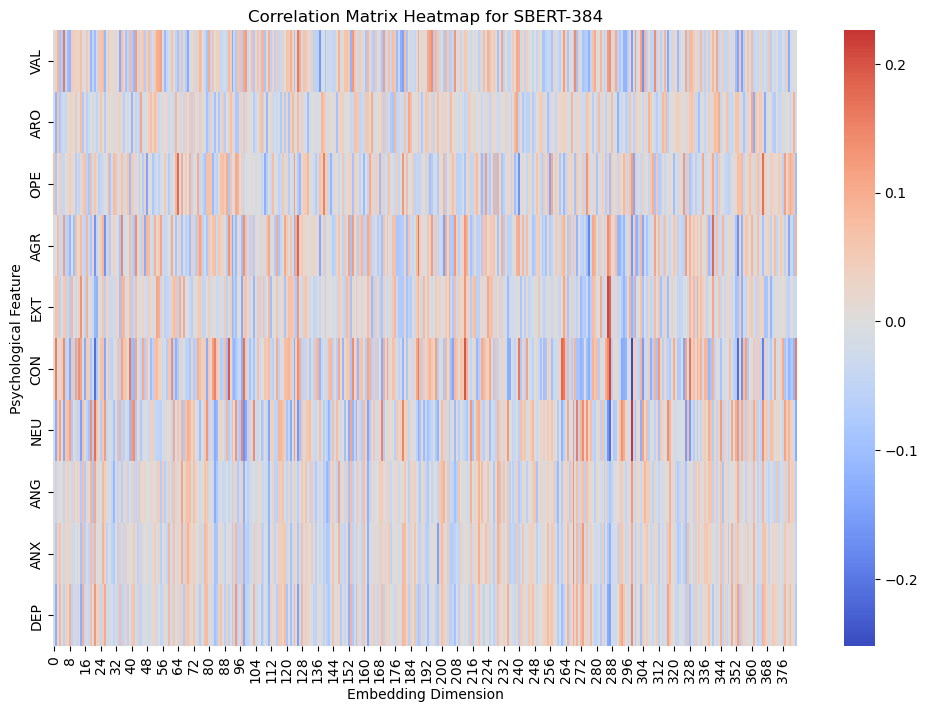}
\caption{Pearson r correlation heatmap of SBERT-384's mean embedding. This visual displays the correlations of SBERT's 384 dimensions with each of the 10 \psychemb dimensions.}
\label{fig:heatmap}
\end{figure}

During the training of WhiSPA, we experimented with identifying which dimensions of the teacher-model, SBERT, have the lowest correlations with \psychemb dimensions to replace those dimensions which is depicted in \autoref{fig:heatmap}. We decided that this approach may lead to statistical biases when training, and so we naively replaced the first 10 dimensions. One should note that the set of 10 dimensions to replace in SBERT can be chosen arbitrarily since our study experimented with this.

\subsection{Annotations}
Please note that the annotators were expert psychologists and co-authors.

The documentation accompanying the iHiTOP interview dataset was utilized to report the coverage of its domains, demographic information, and other relevant details. The dataset’s focus on structured psychological interviews and its linguistic properties were described in the paper to contextualize its relevance to this research. This information was presented to ensure transparency and reproducibility. 
The WTC dataset assessed PTSD symptom severity and related constructs, including anxiety and depression, using English-language data from WTC emergency responders in the Stony Brook Health Program.
The WTC dataset assessed PTSD symptom severity and related constructs, including anxiety and depression, using English-language data from WTC emergency responders.
The development dataset included 1,437 participants, and the prospective dataset included 346, with a mean age of 58 years, predominantly male (93\% and 91\%, respectively) and white (54\% and 49\%).
The analysis emphasized language markers of stress, anxiety, and trauma while reflecting on participants’ experiences of 9/11. Ethical safeguards, including IRB approval, informed consent, and automated anonymization, ensured compliance. While comprehensive in its linguistic and demographic scope, the study was limited to English speakers and WTC responders, constraining generalizability.

\begin{figure}[h!]
    \centering
    \includegraphics[width=\columnwidth]{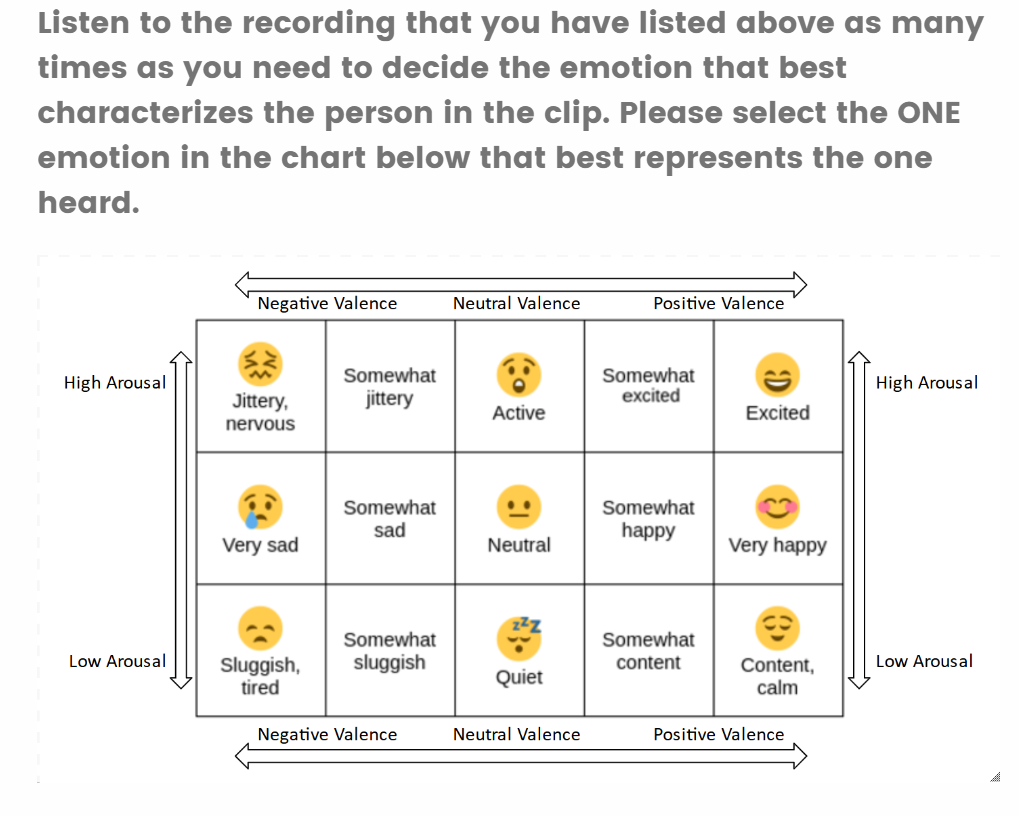}
    \caption{Annotator's affective circumplex visual grid for the task of manually annotating speech segments.}
    \label{fig:affective-circumplex}
\end{figure}

\subsection{Quantitative Analysis}
\label{app:quant_analysis}
\psychemb's lower correlations in \autoref{fig:model_bars} should not be mistaken for poor performance.
With only 10 dimensions, \psychemb representations achieve a staggering $24$ and $22$ Pearson points on \textbf{INT} and \textbf{DIS} respectively, emphasizing its validity as the psychological teacher.
WhiSPA's consistent improvement over the audio models is attributed to the semantic and psychological dimensions that SBERT and \psychemb offer.
Notably, WhiSPA exemplifies drastic improvements in prediction accuracy for all outcomes compared to Whisper.

While WhiSPA demonstrates substantial advancements, surpassing even its text-based LM teacher, SBERT-1024, it remains inherently constrained by the representational capacity of the teacher model.
If the teacher's capabilities are limited, these deficiencies inevitably carry over to the student, even after distillation.
This is evident in the \textbf{ARO} column, where arousal —- an affective dimension -— is more accurately conveyed through acoustic cues.
However, WhiSPA struggles to capture and preserve the acoustic information, instead predominantly aligning with the semantic representations provided by SBERT, thus limiting its ability to fully represent the nuanced affective content inherent in speech.

Beyond demonstrating superior alignment with established PTSD markers, \autoref{tab_app:quant_ngrams} highlights WhiSPA’s enhanced sensitivity to psychologically meaningful language patterns.
\autoref{tab_app:quant_ngrams_pos} shows that n-grams reflecting personal experiences, self-referential content (e.g., first-person pronouns), and negative affective states correlate more strongly with WhiSPA’s predictions than with those of Whisper.
WhiSPA appears better attuned to indicators of psychological distress, anxiety, and trauma symptoms—-an advantage likely stemming from the contrastive alignment objective with text-based representations.
The model’s capacity to detect nuanced emotional and cognitive expressions in spoken language is further supported by its higher effect sizes on known PTSD‐relevant n-grams, underscoring that semantically oriented embeddings can bolster the recognition of clinically significant markers in audio data.

Meanwhile, \autoref{tab_app:quant_ngrams_neg} points to a distinctive negative association between WhiSPA’s predicted severity scores and n-grams referencing positive affect or social relationships.
This result suggests that the same semantically focused latent space that amplifies negative or distress‐related terms also filters out language tied to more adaptive or supportive experiences.
In practical terms, such an effect could be advantageous for screening or early detection: positive affect or relational talk might serve as a buffer or resilience indicator, thereby inversely correlating with predicted symptom severity.
Taken together, these findings highlight the unique strength of WhiSPAs in capturing a wide spectrum of psychologically relevant linguistic markers, surpassing the granularity offered by audio models alone.

\begin{table*}[b]
    \centering
    \begin{subtable}{0.48\textwidth}
        \centering
        \begin{tabular}{|l|c|c|}
        \hline
            \textbf{n-gram} & \textbf{r (WhiSPA)} & \textbf{r (Whisper)} \\ 
        \hline
            me & 0.261 & 0.211 \\ 
            ptsd & 0.226 & 0.126 \\ 
            mental & 0.200 & 0.076 \\
            because & 0.195 & 0.190 \\
            therapist & 0.188 & 0.088 \\
            anxiety & 0.187 & 0.075 \\ 
            my therapist & 0.175 & 0.089 \\
            my mental health & 0.167 & 0.072 \\
            stress & 0.165 & 0.055 \\ 
            want & 0.161 & 0.098 \\ 
            through this & 0.160 & 0.082 \\ 
            pain & 0.158 & 0.171 \\ 
            body & 0.156 & 0.105 \\ 
            this & 0.155 & 0.113 \\ 
            mental health , & 0.152 & 0.051 \\ 
            i had no & 0.151 & 0.135 \\ 
            depression & 0.148 & 0.101 \\ 
            shit & 0.147 & 0.148 \\ 
            but i can't & 0.145 & 0.041 \\
            flashbacks & 0.144 & 0.113 \\ 
        \hline
        \end{tabular}
        \subcaption{}
        \label{tab_app:quant_ngrams_pos}
    \end{subtable}
    \hfill
    \begin{subtable}{0.48\textwidth}
        \centering
        \begin{tabular}{|l|c|c|}
        \hline
            \textbf{n-gram} & \textbf{r (WhiSPA)} & \textbf{r (Whisper)} \\ 
        \hline
            family & -0.264 & -0.200 \\ 
            will be & -0.201 & -0.108 \\
            college & -0.199 & -0.099 \\
            we've & -0.190 & -0.155 \\ 
            will & -0.182 & -0.065 \\ 
            wife & -0.180 & -0.068 \\ 
            pretty & -0.176 & -0.161 \\ 
            as & -0.172 & -0.127 \\ 
            good & -0.170 & -0.170 \\ 
            hopefully & -0.167 & -0.155 \\
            my wife & -0.165 & -0.070 \\ 
            graduated from & -0.163 & -0.028 \\ 
            would & -0.159 & -0.117 \\ 
            able & -0.154 & -0.051 \\ 
            i would say & -0.153 & -0.098 \\ 
            able to & -0.153 & -0.054 \\ 
            kids will & -0.153 & -0.102 \\ 
            would say & -0.152 & -0.100 \\ 
            vacations & -0.151 & -0.152 \\ 
            lucky & -0.150 & -0.101 \\ 
        \hline
        \end{tabular}
        \subcaption{}
        \label{tab_app:quant_ngrams_neg}
    \end{subtable}
    \caption{(a) Top positively correlated N-grams with WhiSPA prediction for PCL scores on the WTC dataset and the corresponding correlations with Whisper predictions. (b) Top negatively correlated N-grams with WhiSPA prediction for PCL scores on the WTC dataset and the corresponding correlations with Whisper predictions. All correlations are statistically significant (p<.05; Benjamini Hochberg corrected).}
    \label{tab_app:quant_ngrams}
\end{table*}

\end{document}